# Looking for Lurkers: Objects Co-orbital with Earth as SETI Observables

James Benford

*Microwave Sciences, 1041 Los Arabis Lane, Lafayette, CA 94549 USA*

jimbenford@gmail.com

**Abstract**

A recently discovered group of nearby co-orbital objects is an attractive location for extraterrestrial intelligence (ETI) to locate a probe to observe Earth while not being easily seen. These near-Earth objects provide an ideal way to watch our world from a secure natural object. That provides resources an ETI might need: materials, a firm anchor, concealment. These have been little studied by astronomy and not at all by SETI or planetary radar observations. I describe these objects found thus far and propose both passive and active observations of them as possible sites for ET probes.

## 1. Introduction

Alien astronomy at our present technical level may have detected our biosphere many millennia ago. Perhaps one or more such alien civilization was drawn in recently, by radio signals emanating from our world. Or maybe it has resided in our solar system for centuries, millennia or longer. Long-lived robotic lurkers could have been sent to observe Earth long ago. If properly powered, and capable of self-repair (von Neumann probes), they could report science and intelligence back to their origin over very long time scales. Long-lived alien societies may do this to gather science for the larger communicating societies in our galaxy.

I will call such a probe a 'Lurker', a hidden observing probe which may respond to an intentional signal and may not, depending on unknown alien motivations. (Here Lurkers are assumed to be robotic.)

Observing 'nearer-Earth objects' would explore the possibility that there are nearby 'exotic' probes that we could discover or excite.

## 1. Lurker History

Bracewell first advanced the *sentinel hypothesis*: that if advanced alien civilizations exist they might place AI monitoring devices on or near the worlds of other evolving species to track their progress. Such a robotic sentinel might establish contact with a developing race once that race had reached a certain technological threshold, such as large-scale radio communication or interplanetary flight [1].

A probe located nearby could bide its time while our civilization developed technology that could find it, and, once contacted, could undertake a conversation in real time. Meanwhile, it could have been routinely reporting back on our biosphere and civilization for long eras.

In 1974 Lunan hypothesized that a Bracewell probe was responsible for long-delayed echoes of early radio transmissions which were observed in the 1920s [2]. These delays were later found to be better explained as propagation phenomena in the earth magnetosphere. Such magnetospheric ducting is the best understood mechanism for long delayed echoes [3]. Papagiannis suggested searching for Lurkers in the asteroid belt [4].

In David Brin's science-fiction novel *Existence*, Lurkers of several types and generations are residing in Earth orbit and the asteroids, where they have been for millions of years [5]. He also authored an "An Open Letter to Alien Lurkers" [6]. John Gertz made a case for probes instead of beacons for interstellar communicating and reviewed where they may be in our Solar System [7].

## 2. Co-orbital Objects

Looking for such *Bracewell probes* offers a number of advantages over traditional SETI, which is listening to the stars. A promising location to search for Lurkers lies among the *co-orbital objects*, which approach Earth very closely annually at distances much shorter than anything except the moon. They have the same orbital period as Earth. These objects could be natural, such as asteroids which have wandered into this type of orbit for a period of centuries to millennia. Or the objects could be artificial in part or in their entirety. Artificial constructs could be noticed by spectroscopy in the visible or near infrared, unless they are buried.

Some objects have unusual *horseshoe orbits* that are co-orbital with Earth. Sometimes these horseshoe objects temporarily become *quasi-satellites* for a few centuries, before returning to their earlier status (Figure 1). Both Earth and Venus have quasi-satellites, as well as most of the outer planets. Venus may have Trojans [8].

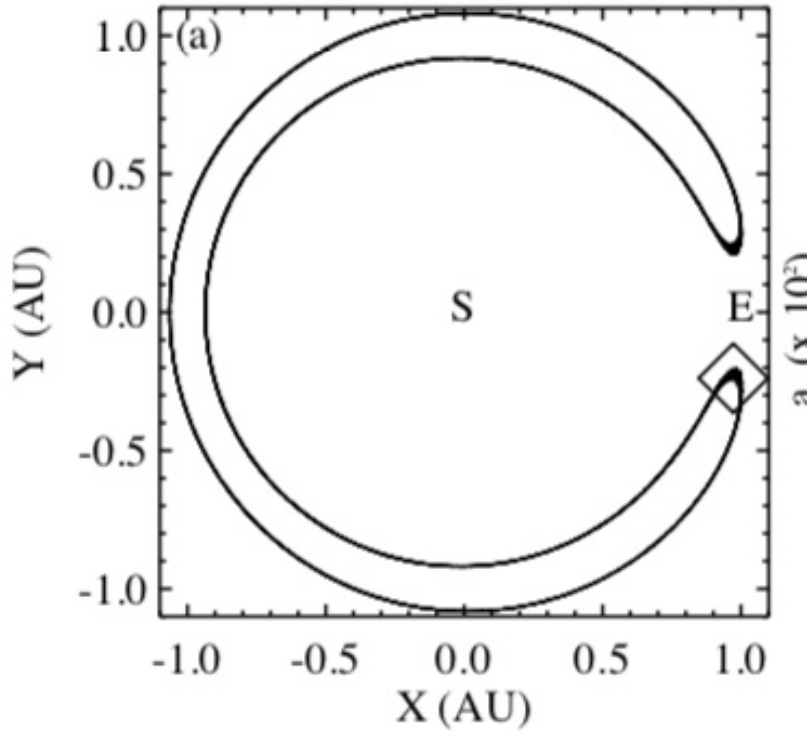

Fig. 1 Horseshoe orbit of 2010 SO16. E is location of Earth, S is the Sun, where the reference frame co-rotates with the Earth's orbit.

A *quasi-satellite* is an object in a 1:1 orbital resonance with a planet. This means the object stays close to that planet over many orbital periods. A quasi-satellite's period around the Sun is the same as the planet, but has a greater eccentricity. When viewed from the perspective of the planet, the quasi-satellite will appear to travel in an oblong retrograde loop around the planet (Figure 2).

They are not true satellites; they are outside the Hill sphere. (An astronomical body's Hill sphere is the region in which it dominates the attraction of satellites, meaning the Hill sphere approximates the gravitational sphere of influence on a small body by a more massive body.)

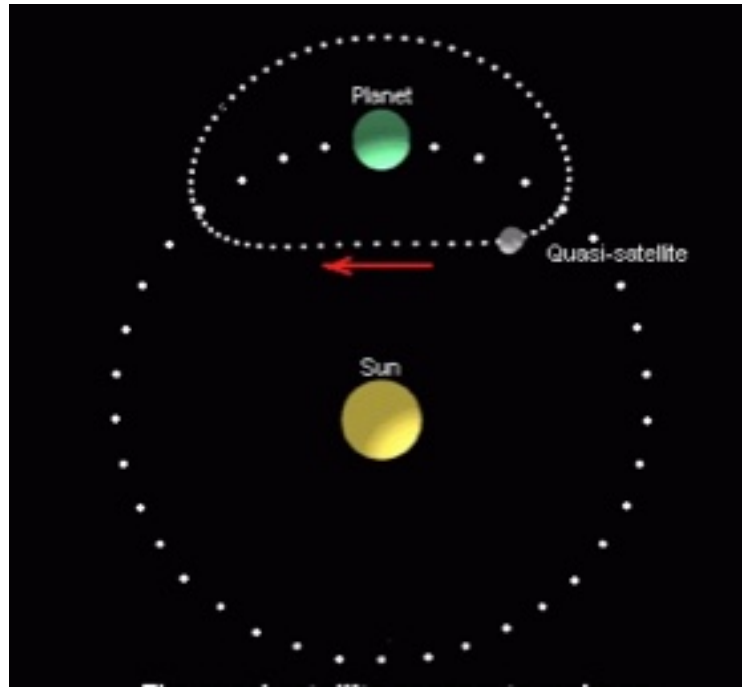

Fig. 2 Quasi–satellite appears to have an oblong orbit as seen from the planet.

Earth's quasi–satellites have been discovered in the last decade; it's likely that quite a few more will be found. There are three configurations: 'horseshoe' and 'tadpole' and 'quasi-satellite' orbits (Figure 3). Their orbits are stable for centuries or much longer. They are <u>not</u> Earth-crossing asteroids (Apollo asteroids), which follow far larger orbits and spend most of their time far from Earth.

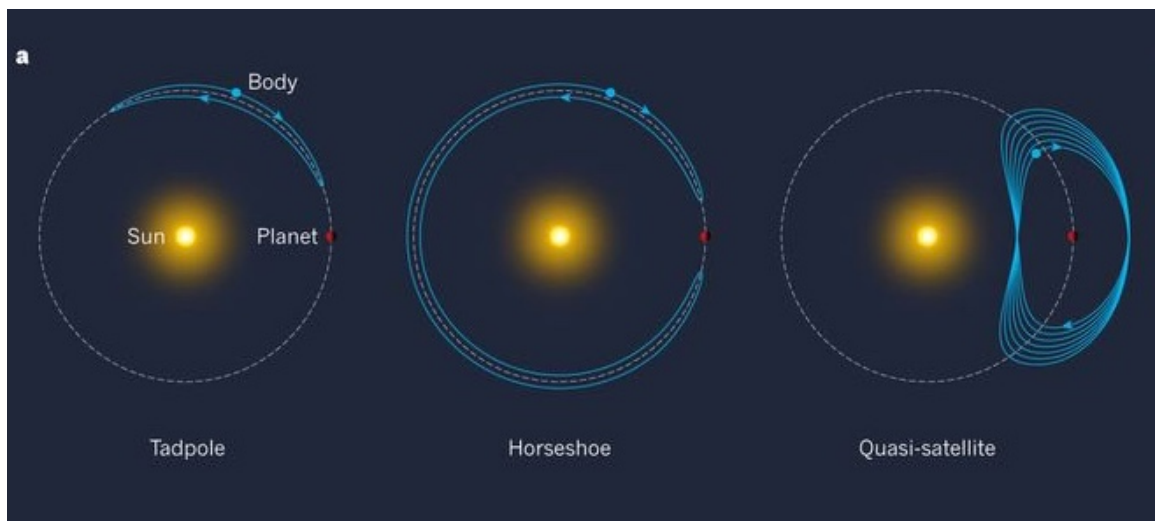

Figure 3 Three types of co-orbital orbits.

Examples of co-orbitals:

- 'Second moon of Earth' Cruithne (3753), estimated size 2 km, closest approach to Earth 0.080 AU = 12 Mkm. [9]. Origin unknown, as it experienced close encounter with Mars 2500 years ago. It is pronounced kru-EEN-ya.
- 'Earth Trojan' (2010 TK$_7$) size 0.3-0.5 km, oscillating about Sun-Earth Lagrangian point L$_4$, so 0.81 to 1.19 AU from Sun, 21 degrees from eliptic in 'tadpole' orbit, closest approach 0.13 AU = 20 Mkm, 50 times Moon distance [10].
- 'Earth's Constant Companion' 2016 HO$_3$ (469219, now called Kamoʻoalewa) is 40-100 m in diameter. It is currently the smallest, closest, and most stable (known) quasi-satellite of Earth, discovered in 2016. Minimum distance is 0.0348 AU = 5.2 Mkm [11].
- Other quasi-satellites are: (164207), 2015 SO$_2$, (227810) 2006 FV$_{35}$, 2013 LX$_{28}$, 2014 OL$_{339}$, 2010 SO$_{16}$ and (469219).

Figures 4-6 show aspects of their orbits. Figure 7 shows the orbits of several of these objects relative to Earth. Table 1 gives some of the key parameters of several of them. Other quasi-satellites are: (164207), (227810) 2006 FV$_{35}$, 2013 LX$_{28}$, 2014 OL$_{339}$, 2010 SO$_{16}$ and (469219).

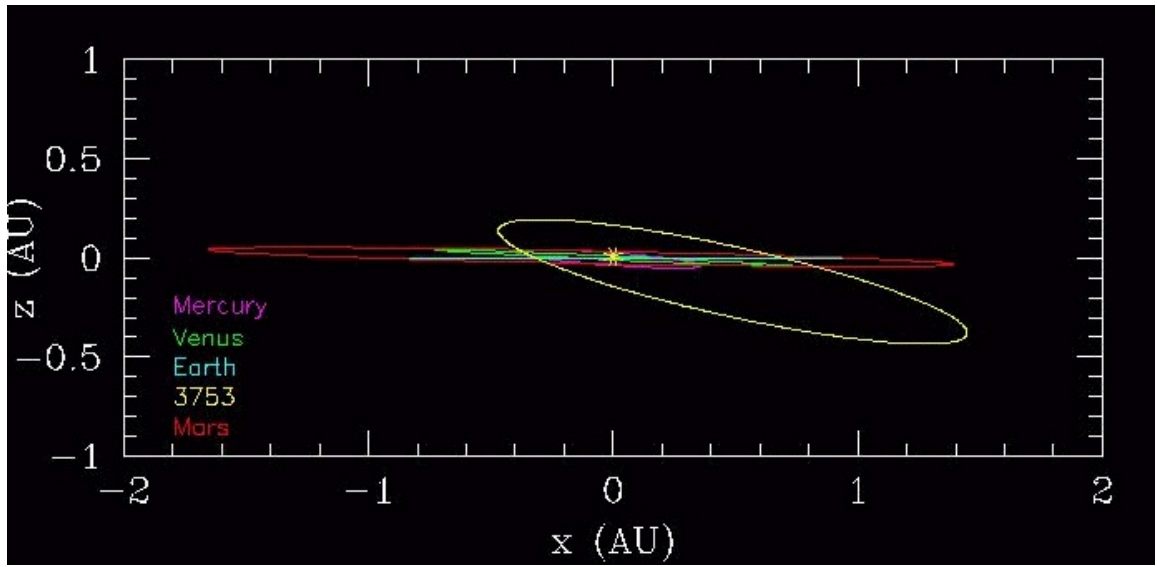

Fig. 4 Cruithne orbit is tilted to the ecliptic.

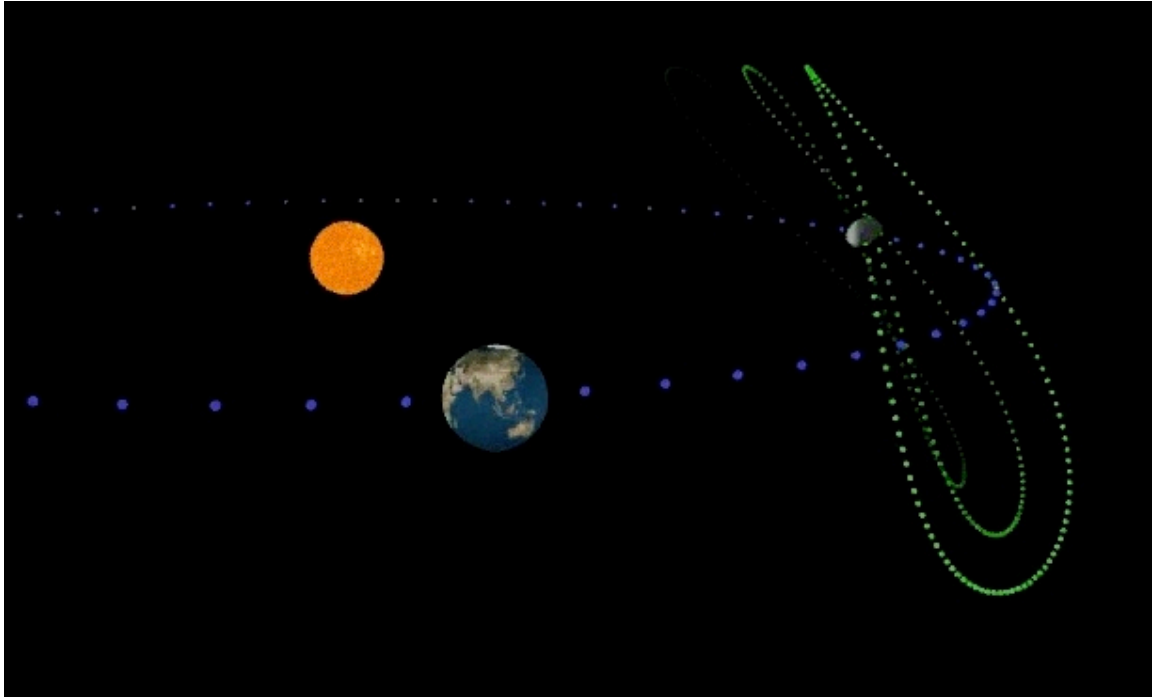

Fig. 5 Earth Trojan orbit [2010 TK$_7$].

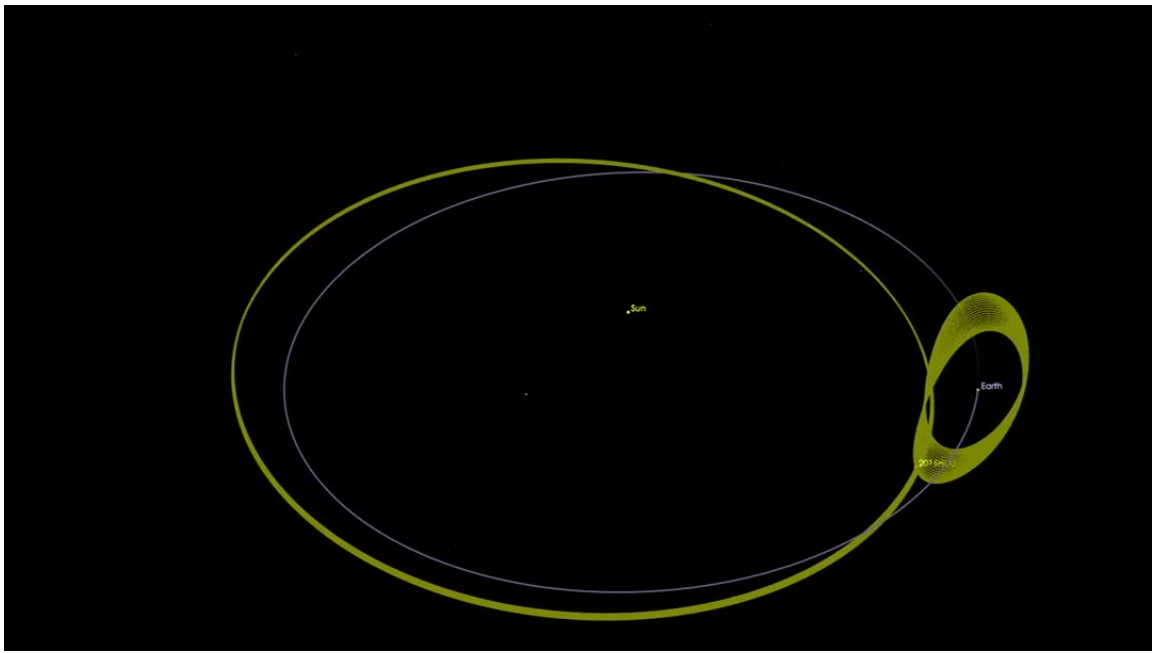

Fig. 6 Orbit of 2016 HO$_3$ around Earth.

Figure 7 shows the orbits of several of these objects relative to Earth. Table 1 gives some of the key parameters of several of them.

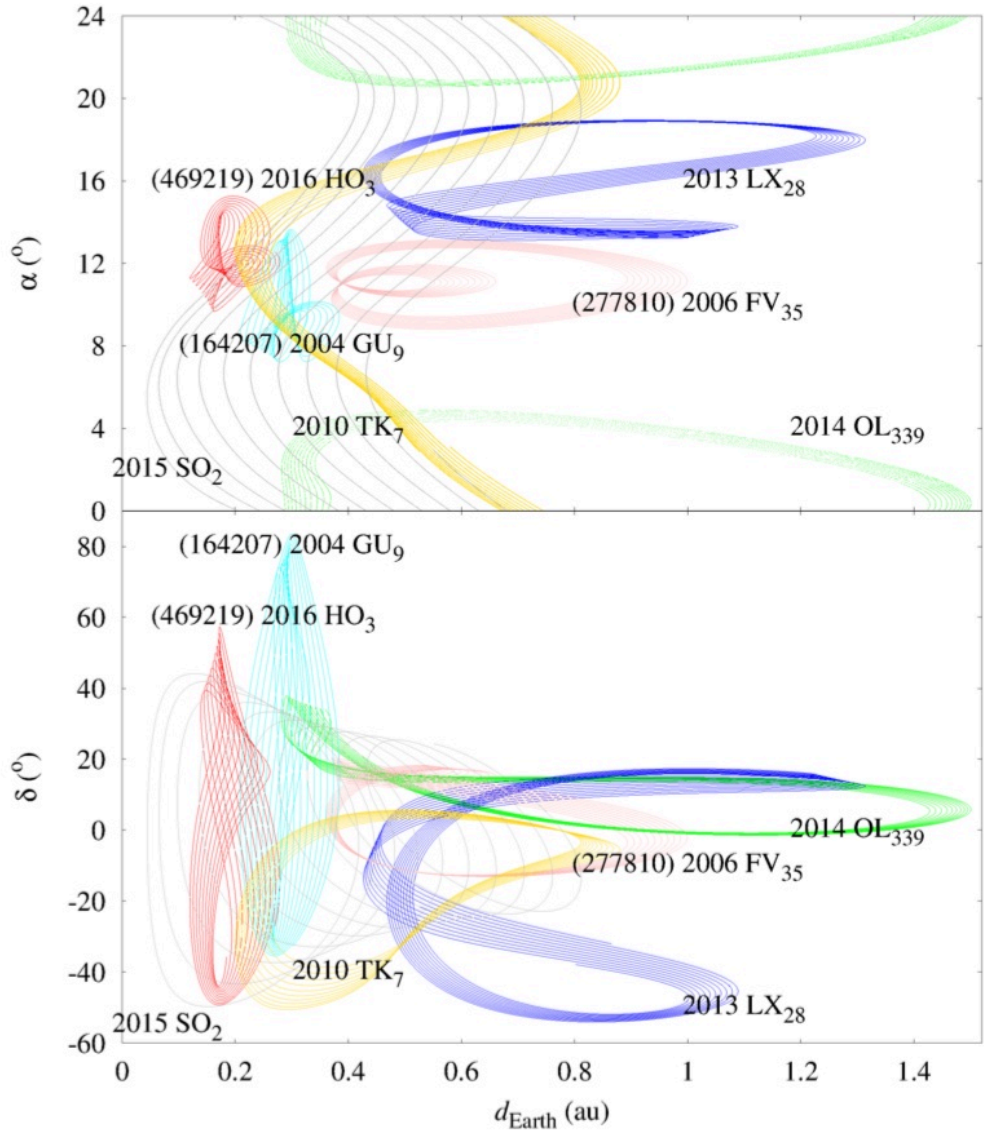

Fig. 7 Orbits of several of these objects relative to Earth in geometric equatorial coordinates. The Y-label of the right ascension (alpha), the units are hours (h) not degrees (o)[11].

| Name | Orbit Type | Size | Closest Approach | Albedo | Magnitude |
|---|---|---|---|---|---|
| 3753 Cruithne | Horseshoe | 2 km | 12 Mkm | 0.365 | 15.6 |
| 'Earth Trojan' 2010TK$_9$ | Tadpole | 0.4 km | 1.5 Mkm | 0.06 | 20.8 |
| 'Earth's Constant Companion' 2016 HO$_3$ | Tadpole | 0.04 km | 5.2 Mkm | 0.25 | 24.3 |
| 2006 FV$_{35}$ | Horseshoe | 0.14-0.32 km | 15 Mkm | 0.22 | 21.1 |
| 2014 OL$_{339}$ | Tadpole | 0.17 km | 2.7 Mkm | (0.19) | 22.9 |
| 2004 GU$_9$ | Tadpole | 0.22 km | 0.0004 Mkm | 0.22 | 21.1 |
| 2010 SO$_{16}$ | Tadpole | 0.36 km | 0.029 Mkm | 0.084 | 20.5 |
| 2015 SO$_2$ | Horseshoe | 0.05-0.11 km | 0.019 Mkm | 0.108 | 23.9 |

- 

Table 1 Parameters of several co-orbital objects. Albedos in parenthesis are estimates used with the observed magnitude to derive the diameter.

### 3. Stability of Co-orbitals

The long term stability of these objects is an area of substantial study [12-15]. A large number of horseshoe co-orbitals of Earth, quasi-satellites, Trojans and horseshoe orbits, appear to be long-term stable. Morais and Morbidelli, using models of main asteroid belt sources providing the co-orbitals and their subsequent motions, conclude that the mean lifetime for them to maintain resonance with Earth id 0.33 million years [16].

There are interesting examples: Other examples near Earth are:

- Cruithne may leave its orbit 5000 years from now by interaction with the giant planets.

- 2010 TK9 orbital parameters indicate that most probably the asteroid became a Trojan 1.800 years ago and will remain there for ~15,000 years, when it will jump to a horseshoe orbit.

- 2016 HO$_3$ has been a stable quasi-satellite of Earth for almost a century, and it will continue to follow this pattern as Earth's closest companion for centuries to come. It never wanders farther away than about 100 times the distance of the moon or comes closer than 38 times that. 2016 HO$_3$ swam into its orbit close to Earth only a century ago, a remarkable coincidence in cosmic time.

- 2006 FV$_{35}$ has occupied its orbit state for about 100,000 years and will stay in this orbit for about 800 more years.

- 2014 OL339 became a quasi-satellite at least 775 years ago and will stop being that 165 years from now after a "close" encounter with Earth at ~0.13 AU. This quasi-satellite episode will have had a duration of <2,500 years. It is the most unstable of the known Earth quasi-satellites.

- Two close quasi-satellites, 2015 $SO_2$ and 2016 $HO_3$, have the unusual coincidental feature of orbits that have been almost identical for many years and approach Earth at the same time every year [11].

**4. SETI Searches of Co-orbitals**

Several approaches to study these objects, starting with passive observations:

1) Plan a multiyear program of observations by radio and optical telescopes and planetary radars around the world. Central to the project would be optical telescopes, such as the Lick Observatory and other platforms participating in the Breakthrough Listen project, to discern their size, shape, rotation periods, and optical properties, such as spectra. We would need to discern their optical spectra out to at least J-band (to 1.2 µ). A Bracewell probe could also give specular reflection of sunlight from its artificial surfaces [17].

2) Conduct passive SETI observations of these nearer-Earth objects in the microwave, infrared and optical.

3) Use active planetary radar to investigate the properties of these objects. These objects have not been pinged or imaged by any planetary radar as yet. Recent developments in planetary radars have shown they can detect the presence and trajectories of spacecraft in lunar orbit, even though their size is a few meters [18]. The 'glinty' reflections from spacecraft, large rapid changes in signal-to-noise ratio, would be an indicator of an artificial object [18]. Whether these radars are sensitive or powerful enough to get a return signal for imaging from any of the presently known co-orbital objects requires analysis. I estimate that 2016 $HO_3$ is detectable with a signal-to-noise ratio>100. In any case, they can "ping" the objects, meaning that a signal reaches them, but the return signal may be too weak to detect at Earth. If there is an ET probe there, it might sense that it had been noticed by us. We could imprint a message on these signals.

4) Conduct *active* simultaneous planetary radar 'painting' and SETI listening of these objects. This would be 'Active SETI', *which could solicit a response from a hypothetical probe.* This does not incur the objections to sending interstellar messages, messaging to ETI (METI), because any such alien lurkers would already know we are here. Of course, this is at very short range compared to the interstellar ambitions of METI enthusiasts. We presume that Lurkers already know that we have radar, but might not respond to a single simple radar painting such as we have done to many asteroids. If we want to send a message, as Paul Davies suggested for the Lagrange points in 2010, how would a signal be designed to elicit such a response [4, 19]? This is the basic question of METI. Figuring out this near-term possible use of METI can drive discussion and research ideas in this field because it's a concrete problem: what message would draw them out of their passive state to interact with us? One straightforward message to send would be a photograph of the object we are sending the message to. Taking the highest resolution pictures of it as it rotates would simply say "We see you."

5) Launch robotic probes and manned missions to conduct inspections, take samples. For example, 2016 HO$_3$ at close approach has a relative velocity of 3-5 km/sec, so is within present capability. 2006 FV$_{35}$ requires ~12 km/sec to rendezvous [20].

Perhaps probes are waiting on these objects, listening to us and waiting for us to find them. They may remain silent and simply report back to wherever they communicate to.

## 5. *Not* Favored For Lurking Near Earth

There are several other regions that a Lurker might locate. Here I describe some that are less promising than the co-orbitals.

- Earth orbit: Orbits very near Earth are not stable over the long term due to drag. Geosynchronous orbits are stable. They are quite observable by optical and radar means, but all that have been seen are ours.

- Lagrangian points : The Earth-Moon L$_4$ and L$_5$ Lagrangian points contain only interplanetary dust in what are called *Kordylewski clouds*. At least one asteroid, 2010 TK7, detected in October 2010, oscillates about the Sun–Earth L$_4$ Lagrangian point (~60 degrees ahead of Earth. See section 2.) There are probably many others at the Sun–Earth Lagrangian points [21].

- Moon: The moon of course is the nearest object. From orbiters we have photographs of almost all the moon at resolutions of ~1 meter, and teams of people looking at them carefully. Nothing has been seen except our own artifacts we sent there. So then one would have to presume that observers would have built in a very disguised way. (As in *2001*, where the Monolith was "deliberately buried", to ensure that only a civilization capable of spaceflight would be able to discover it.) Any point on the Moon is in shadow half the time, so is cold and has to have storage to operate through the night if it is solar powered. Moreover, if a probe is going to respond to us, it would surely have done so by now, responding the many transmissions that been made to the moon for communications with our orbiters and landers.

- <u>Earth-crossing asteroids</u> (Apollo asteroids), which follow far larger orbits and spend most of their time far from Earth, so are not as useful for Earth-observing as are the co-orbitals.

- <u>Asteroids</u>: Some of the swarm of asteroids in the Belt might be used by Lurkers, but have many drawbacks as a location: Of course, asteroids are much further away. And at several AU, the solar flux is reduced by a factor of 4-10. Therefore solar panels would have to be a very large. Nuclear powered systems would be preferred. (This is another reason to lurk near Earth: higher solar flux.) The asteroids are very cold for the same reason. That means that mechanical systems and electrical systems, as well as lubricants and everything associated with chemistry would be far more difficult to keep working over long times. However, there are potential reasons to lurk there: (i) availability of metal and volatile-rich materials, (ii) safety from orbital perturbations over much longer time scales … tens or hundreds of millions rather than mere tens of thousands of years, (iii) opportunity to evaluate a new spacefaring culture in safety from discovery for longer periods.

## 6. Discussion

General Observations:

- An overall reason to look closer to Earth is that we haven't seen anything in the rest of the solar system. And we haven't seen anything communicating from the nearby stars out to about 100 light years.
- The basic fact is we do not know alien logic, alien instinct, alien intention or anything else about them.
- The civilization that sent the probe may be very long-lived, meaning that individuals may live much longer than we do or that they are in fact AI's. If properly powered, and capable of self-repair (von Neumann probes), they could report science and intelligence back to their origin over very long timescales. Therefore, looking locally near Earth not only explores a new space, but also Deep Time [22, 23].
- To study these nearby objects changes the means of inspection from listening to the stars to astronomical diagnostics such as imaging and spectroscopy. This approach involves techniques well developed for the study of asteroids and planets. That means using other technologies and other institutions to pursue this search.
- This possibility allows a local test of messaging (METI) prospects. We can construct messages and try them against nearby objects, thus circumventing arguments that we might be noticed by or encourage hostile forces in the stars. The SETI and METI landscape is thus transformed into a local experiment. Many of the arguments against interstellar METI (e.g. drawing attention to ourselves) seem less compelling when it comes to attempted messaging of potential Lurker sites that are so close by, since such nearby Lurker will already know of us.
- Civilizations would not need to be very close to the solar system to send a probe here in the last few centuries. Here are some numbers: Suppose a probe can be launched at 1/10th speed light, which is certainly something we have figured out how to do conceptually by either beam-driven sails (photon beams, neutral particle beams) or nuclear fusion rockets. Traveling over 300 years, such probes could reach us from 30 light years out. About 3000 main sequence stars lie within 100 light years, so plenty of stars that might have civilization.
- For societies with time horizons beyond a century-long payoff, lower speeds like 0.01 c demand a millennium to take up residence and study Earth. Worlds that take even longer perspectives can use lower speeds and so lower their initial costs. Note that this implies that by searching locally, we extend our quest over both long distances in space and great spans in time, generalizing the entire SETI approach. Even alien societies that have ceased their interstellar interests or are even extinct can still tell us something, through their ancient probes.

What have we to lose by checking out these objects? Certainly resources such as time on telescopes, radio and optical. But we would be studying newly found objects, which could well be interesting astronomy. Nobody has really looked at these co-orbitals, other than orbital calculations and faint images. We know almost nothing about them.

## 7. Conclusion

Co-orbitals are attractive targets for SETI searches because of their proximity. We should move forthrightly toward observing them, both by observing them in the electromagnetic spectrum and planetary radar, as well as visiting them with probes. The mast attractive target is 'Earth's Constant Companion' 2016 $HO_3$, the smallest, closest, and most stable (known) quasi-satellite of Earth. Getting there from Earth orbit requires a delta-v of about 4.5 km/sec. It approaches Earth annually, in October. China has announced they are going to send a probe to 2016 $HO_3$ [24].

**Acknowledgements**

I gratefully acknowledge funding by the Breakthrough Foundation.

Thanks to Dominic Benford, Gregory Benford, David Brin, Keith Cooper, Paul Davies, Carlos de la Fuente Marcos, John Gertz, Paul Gilster, Al Jackson, Tom Kuiper, Joe Lazio, Geoff Marcy, Andrew Siemion and Marty Slade for comments.**References**

1. R. N. Bracewell, "Communications from Superior Galactic Communities," *Nature*, 186, pg. 670, 1960. Reprinted in A.G. Cameron ed., *Interstellar Communication*, W. A. Benjamin, Inc., New York, pg. 243, 1963.
2. D. Lunan, *Man in the Stars*, London Souvenir Press 1974. Published in the US as *Interstellar Contact*, Henry Regnery Chicago, 1975.
3. R. J. Vidmar and F. W. Crawford, "Long-delayed radio echoes: Mechanisms and observations," Journal of Geophysical Research, 90, pg. 1532, 1985.
4. M. D.Papagiannis, "Are We Alone, or Could They be in the Asteroid Belt?", *Quarterly Journal Royal Astronomical Society*, 19, pg. 277, 1978.
5. D. Brin, "*Existence*", Tor Books, New York, 2012.
6. D. Brin, "An Open Letter to Alien Lurkers", Invitation to ETI, http://www.ieti.org/articles.brin.htm. (Posted March 1999, last Accessed 9 March 2019).
7. J. Gertz, "ET Probes: Looking Here as Well as There", *JBIS* 69, pg. 88, 2016.
8. P. Pokorný and M.Kuchner, "Co-orbital Asteroids as the Source of Venus's Zodiacal Dust Ring", *ApJ Lett.,* 843, L16, 2019.
9. P. A. Wiegert, K. A. Innanen and S. Mikkola, "An asteroidal companion to the Earth", *Nature*, 387, pg. 685,1997.
10. A. Mainzer, J. Bauer, T. Grav and 32 coauthors, "Preliminary Results from NEOWISE", *ApJ*, 731, 53, 2011.
11. C. de la Fuente Marcos and R. de la Fuente Marcos, "Asteroid (469219) 2016 HO3, the smallest and closest Earth quasi-satellite", *Monthly Notices of the RAS,* 462, 2016.
12. M, Ćuk, D. Hamilton and M. Holman, "Long-term stability of horseshoe orbits", *Monthly Notices Royal Astronomical Society,* 426, pp. 3051, 2012.
13. Zhou, Lei; Xu, Yang-Bo; Zhou, Li-Yong; Dvorak, Rudolf; Li, Jian, "Orbital stability of Earth Trojans", *Astronomy & Astrophysics*, 622, 14 , 2019.
14. F. Marzari, H. Scholl, "Long term stability of Earth Trojans", *Celestial Mechanics and Dynamical Astronomy*, 117, pp.91, 2013
15. P. Wiegert, K. Innanen and S. Mikkola, "The Orbital Evolution of a Near-Earth Asteroid 3753", *Astronomical Journal,* 115, 1998.
16. M. Morais and A. Morbidelli, 'The Population of Near-Earth Asteroids in Coorbital Motion with the Earth", Icarus 160, pg. 1, 2002.
17. B. C. Lacki, "A Shiny New Method For SETI: Specular Reflections From Interplanetary Artifacts", arXiv: astro-ph/1905.05839v1, 14 March 2019.
18. Brozovic et al., "Radar Observations of Spacecraft in Lunar Orbit", ISTS-2017.
19. P. Davies, *The Eerie Silence,* pg. 107, Houghton Mifflin Harcourt, 2010.13